\documentclass[%
 reprint,
 amsmath,amssymb,
 aps,
]{revtex4-2}

\usepackage{xcolor}
\usepackage{hyperref}
\hypersetup{
    colorlinks,
    linkcolor={red!80!black},
    citecolor={blue!80!black},
    urlcolor={blue!80!black}
}
\usepackage{comment}
\usepackage{units}
\usepackage{amsmath}
\usepackage{graphicx}
\usepackage{dcolumn}% Align table columns on decimal point
\usepackage{bm}
\usepackage [english]{babel}
\usepackage [autostyle, english = american]{csquotes}
\MakeOuterQuote{"}
\graphicspath{{./}}
\begin{document}

\title{Self-lensing flares from black hole binaries III: \\ general-relativistic ray tracing of circumbinary accretion simulations}

\author{Luke Major Krauth$^{1}$}
 \thanks{EMAIL: LMK2202@columbia.edu}
\author{Jordy Davelaar$^{2,3,4}$}
\author{Zoltán Haiman$^{1,2}$}
\author{John Ryan Westernacher-Schneider$^{5}$}
\author{Jonathan Zrake$^{6}$}
\author{Andrew MacFadyen$^{7}$}

\affiliation{
$^{1}$Department of Physics, Columbia University, New York, NY 10027, USA}
\affiliation{
$^{2}$Department of Astronomy, Columbia University, New York, NY 10027, USA}
\affiliation{
$^{3}$Astrophysics Laboratory, Columbia University, 550 W 120th St, New York, NY 10027, USA}
\affiliation{
$^{4}$Center for Computational Astrophysics, Flatiron Institute, 162 Fifth Avenue, New York, NY 10010, USA}
\affiliation{
$^{5}$Leiden Observatory, Leiden University, P.O. Box 9513, 2300 RA Leiden, The Netherlands}
\affiliation{
$^{6}$Department of Physics and Astronomy, Clemson University, Clemson, SC 29634, USA}
\affiliation{
$^{7}$Center for Cosmology and Particle Physics, Physics Department, New York University, New York, NY 10003, USA
}



\begin{abstract}

Self-lensing flares (SLFs) are expected to be produced once or twice per orbit by an accreting massive black hole binary (MBHB), if the eclipsing MBHBs are observed close to edge-on. SLFs can provide valuable electromagnetic (EM) signatures to accompany the  gravitational waves (GWs) detectable by the upcoming Laser Interferometer Space Antenna (LISA). EM follow-ups are crucial for, e.g., sky-localization, and constraining the Hubble constant and the graviton mass. We use high-resolution two-dimensional viscous hydrodynamical simulations of a circumbinary disk (CBD) embedding a MBHB. We then use very high-cadence output of these hydrodynamical simulation inputs for a general-relativistic ray-tracing code to produce synthetic spectra and phase-folded light curves. Our main results show a significant periodic amplification of the flux with the characteristic shape of a sharp flare with a central dip, as the foreground black hole (BH) transits across the minidisk and shadow of the background BH, respectively. These corroborate previous conclusions based on the microlensing approximation and analytical toy models of the emission geometry. We also find that at lower inclinations, without some occlusion of the minidisk emission by the CBD, shocks from quasi-periodic mass-trading between the minidisks can produce bright flares which can mimic SLFs and could hinder their identification.

\end{abstract}
\maketitle

%\keywords{ }

%% cross-references to sections, equations, tables, and figurers.
\section{Introduction}

It is currently believed that massive black holes (BHs) are found in the cores of virtually all galaxies \cite[e.g.][]{Kormendy2013}. When two such galaxies merge, various forces, such as dynamical friction and gravitational interactions with gas and stars, can move the two BHs close enough that they become gravitationally bound, forming a massive black hole binary (MBHB) \cite[e.g.][]{Mayer2013}. Strong tides associated with the galaxy merger are thought to trigger flows of gas toward the galaxy core, which likely form an accretion flow around the binary in the form of a circumbinary disk (CBD) \cite[e.g.][]{begelman1980}. The CBD, and the "minidisks" which form around the component BHs, could provide a source of observable electromagnetic (EM) emission.

Hydrodynamical simulations over recent years have converged on describing several features of these MBHB+CBD systems: even though the binary system is expected to carve out a low-density central cavity which extends to several times its orbital separation, the BHs are not expected to starve, as each component is fueled by narrow streams of gas, which are tidally stripped from the CBD, and lead to the formation of the minidisks around each BH. As such, electromagnetic (EM) signatures can originate from several sources: the CBD, the minidisks, or the streams colliding with each other or the cavity wall~\citep[see, e.g.][and references therein]{westernacher-schneider2022}. The fluctuations in the brightness of the system are often approximately periodic on the binary’s orbital timescale~\citep[see, e.g.][for early works]{artymowicz1996,hayasaki2007,macfadyen2008,Roedig2011,shi2012,noble2012,dorazio2013,farris2014}. Additionally, gas can build up a non-axisymmetric density distribution at the far edge of the cavity, known as a ``lump.'' A flow pattern associated with the lump propagates along the cavity edge, periodically modulating accretion onto the minidisks on several times the orbital timescale, as well as on beat periods between the binary and the lump~\citep[e.g.][]{shi2012,dorazio2013,farris2014,shi2015,noble2021,mignon2023}. The minidisks can also exchange mass, which can introduce periodicity slightly different from the orbital period~\citep{bowen2017, westernacher-schneider2022, WS+2023}. Further periodic signatures are possible through relativistic Doppler boosting~\citep{bode2010,dorazio2015,haiman2017}. These features are not necessarily generic however, and can depend on eccentricity, mass ratio of the BHs, disk temperature, and other factors of the system \citep[e.g.][]{noble2021,westernacher-schneider2022}. As such, amongst the variety of predicted emission mechanisms, discovering a robust signature that can be isolated amongst the noise could be crucial for identifying sources.

Viable candidates for such unique signatures are self-lensing flares (SLFs). As the two BHs align with the line of sight for a nearly edge-on observer, the EM emission from the vicinity of the background BH can be lensed by the steep gravitational potential of the foreground BH, if it is within the Einstein angle of the foreground BH.  
This angle is defined as~\citep[e.g.][]{Gaudi2012}:

\begin{equation}
    \theta_{\rm E} = r_{\rm E}/D =\sqrt{\frac{4GM}{Dc^2}},
\end{equation}
where $r_{\rm E}$ is the Einstein radius, $D^{-1}\equiv D_L^{-1}-D_S^{-1}$, where $D_L$ and $D_S$ are the (luminosity) distances to the lens and the source, respectively, $G$ is Newton's constant, $M$ is the mass of the lens, and $c$ is the speed of light. SLFs can produce a considerable amplification of the observed flux. This signature is expected to happen twice per full orbit of the system (assuming both BHs are bright enough to be detectable), and is sufficiently unique to distinguish a binary from other possible sources.

Because of the bending of light rays around the BH, sight lines passing within $2.6\,r_\mathrm{s}$ (where $r_\mathrm{s} \equiv 2GM/c^2$) of a (non-rotating) BH terminate on the horizon \cite{Falcke2000}. This is known as the BH shadow. A recent prediction by \citep{davelaar2022,davelaar2022b} showed that as the focal point from the foreground BH passes over the shadow of the background BH, a "dip" could be seen at the peak of the amplification during the SLF. Since the shadow of a BH directly corresponds to its mass, the duration of a dip at the peak of the SLF could enable measurement of the BH masses; this can even be done at cosmological distances where spatially resolved observations are infeasible, even with current very long baseline interferometry (VLBI). Identification of SLFs could also help refine the sky localization of binaries observed with LISA or the Vera Rubin Observatory \cite{LSST2021}. Combining the electromagnetic and gravitational wave signatures enables multi-messenger astrophysics analyses, ranging from constraining the mass of the graviton \cite{Haiman2017b}, to independently measuring the Hubble constant \cite{Feeney2019}, to providing new constraints on accretion physics and the co-evolution of galaxies and their central massive BHs (see, e.g., \citep{baker2019} for a brief review and references).

Past studies have investigated SLFs \cite{Hu2020,dorazio2018,Beky2013,haiman2017} using a microlensing approximation (where the BHs are treated like point sources) \cite{Paczynski1986} or ray-tracing analytical models \cite{Ingram2021,davelaar2022,davelaar2022b}. We build on the findings of these studies \cite[specifically][hereafter Paper I and II]{davelaar2022,davelaar2022b} by first simulating accreting MBHBs in two dimensions with the astrophysical gas dynamics code \texttt{Sailfish} \footnote{https://github.com/clemson-cal/sailfish}. We then adapt the output, using it as input for our emission model in a suite of post-processing runs in the general-relativistic ray-tracing code \texttt{RAPTOR} \cite{Bronzwaer2018}, in order to develop a more physically motivated model to test if and how SLFs persist in the presence of realistic accretion variability. This is covered in greater detail in \S~\ref{sec:Numset}. More specifically, in \S~\ref{subsec:HydroSetup} we discuss our hydrodynamics code, numerical schemes, and process. In \S~\ref{subsec:RaptorSetup} we discuss how we have adapted the \texttt{Sailfish} output for \texttt{RAPTOR}, our general-relativistic ray tracing and setup, and our calculation of light curves. In \S~\ref{subsec:Models} we discuss our suite of models. In \S~\ref{subsubsec:Fid} we present our main results on the gas dynamics, spectrum, and EM signatures of our fiducial model. In \S~\ref{subsec:ModCompare} we compare the results for several models with parametrically adjusted disk thickness and warping, for multiple viewing angles. Finally, in \S~\ref{sec:DisCon} we summarise our main conclusions and the implications of this work.

\section{Numerical setup}
\label{sec:Numset}
\subsection{Hydrodynamical setup}
\label{subsec:HydroSetup}

All hydrodynamic simulations were run in two dimensions using the publicly-available graphics processing unit (GPU)-accelerated astrophysical gas dynamics code \texttt{Sailfish}. This section provides a brief summary of the technical aspects of {\tt Sailfish} \citep[for more details, see][]{westernacher-schneider2022}.

{\tt Sailfish} solves the vertically-integrated 
Navier-Stokes equations:

\begin{eqnarray}
    \partial_t \Sigma + \nabla_j \left( \Sigma v^j \right) &=& S_{\Sigma} \label{eq:mass} \\
    \partial_t \left( \Sigma v_i \right) + \nabla_j \left( \Sigma v^j v_i + \delta^j_i \mathcal{P} \right) &=& g_i + \nabla_j \tau^j_i + S_{p, i} \label{eq:mom} \\
    \partial_t E + \nabla_j \left[ \left( E+\mathcal{P} \right) v^j \right] &=& v^jg_j + \nabla_j \left( v^i \tau^j_i \right) \nonumber\\
    &-& \dot{Q} + S_{E} \label{eq:en},
\end{eqnarray}
where $\Sigma$ and $\mathcal{P}$ are the vertically-integrated mass density and pressure, respectively, $v^i$ is the mid-plane horizontal fluid velocity, $E=\Sigma \epsilon + (1/2)\Sigma v^2$ is the vertically-integrated energy density, $\epsilon$ is the specific internal energy at the mid-plane of the disk, $g_i$ is the vertically-integrated gravitational force density, and $\tau^j_i = \Sigma \nu_\mathrm{visc} \left( \nabla_i v^j + \nabla^j v_i - (2/3)\delta^j_i \nabla_k v^k\right)$ is the viscous stress tensor (in a form that is trace-free in a 3-dimensional sense) with zero bulk viscosity, $\nu_\mathrm{visc}$ is the kinematic shear viscosity modelling unresolved turbulence and magnetic fields, $S_{\Sigma}$, $S_{p,i}$, and $S_E$ are mass, momentum, and energy sinks modelling gas consumption by black holes, and $\dot{Q}$ is the local black-body cooling prescription, assuming hydrogen dominates the gas density \citep[see e.g.][]{frank2002}, given by
\begin{eqnarray}
    \Dot{Q} &=& \frac{8}{3} \frac{\sigma} {\kappa\Sigma} \left(\frac{m_p\mathcal{P}}{k_B\Sigma}\right)^4 \nonumber \\
    &=& 2 \sigma T^4_{\rm eff}
    \label{eq:cooling}.
\end{eqnarray}
where $\sigma$ is the Stefan-Boltzmann constant, $\kappa = \unit[0.4]{cm^2 g^{-1}}$ is the opacity due to electron scattering, $m_p$ is the proton mass, and $k_B$ is the Boltzmann constant, and the factor of 2 comes from the fact that the disk cools on both the top and bottom faces. We employ a torque-free sink prescription \citep{dempsey2020, dittmann2021} to model the removal of gas by black holes, with the sink radius equal to the black hole's Schwarzschild radius, $r_{\rm s}$.

The simulations model a geometrically thin and optically thick accretion disk, initialized with near-Keplerian rotation. We use a Shakura-Sunyaev viscosity prescription $\nu_\mathrm{visc} = \alpha c_{\rm s}h$ \citep{shakura1973}, where $c_s = \sqrt{\Gamma\mathcal{P}/\Sigma}$ is the sound speed, $h$ is the disk's half-thickness, and we adopt $\alpha = 0.2$. Following \cite{westernacher-schneider2022}, we use a $\Gamma$-law equation of state resulting in the form $\mathcal{P}=\Sigma\epsilon(\Gamma-1)$, where $\Gamma=5/3$, and we neglect radiation pressure. The disk's half-thickness is given by
\begin{equation}
    h = \frac{\sqrt{\mathcal{P}/\Sigma}}{\tilde{\Omega}} \, ,
    \label{eq:disk-height}
\end{equation}
where $\tilde \Omega$ is a binary analog of the Kepler frequency,
\begin{equation}
    \tilde{\Omega}\equiv\sqrt{GM_1/r_1^3+GM_2/r_2^3} \, ,
    \label{eq:binary-omega}
\end{equation}
and $r_1$ and $r_2$ are the distances to the respective black holes. The characteristic value of $h$ corresponds to a Mach number $\mathcal{M}\sim 30$. If radiation pressure were accounted for, this corresponds to a disk accreting at $\dot{M}=0.19 \,\dot{M}_{\rm Edd}$, where the Eddington accretion rate is $\dot{M}_{\rm Edd}=L_{\rm Edd}/(\eta c^2)$, $L_{\rm Edd}$ is the Eddington luminosity, and $\eta=0.1$ is the radiative efficiency. We initialize the gas profile as an approximate axisymmetric equilibrium:

\begin{align}
    \Sigma &= \Sigma_0\left(\frac{\sqrt{r^2+r_\mathrm{soft}^2}}{a}\right)^{-3/5}\\
    \mathcal{P} &= \mathcal{P}_0\left(\frac{\sqrt{r^2+r_\mathrm{soft}^2}}{a}\right)^{-3/2}\\
    \vec{v} &= \left({\frac{GM}{\sqrt{r^2+r_\mathrm{soft}^2}}}\right)^{1/2}\hat{\phi},
\end{align}

\noindent
where $r$ is the distance from the binary barycenter, $r_{\mathrm{soft}}$ is the gravitational softening length, which models the vertically-integrated gravitational force, and $a$ is the binary semi-major axis. We initialise a central cavity at $r=2a$ by multiplying the initial profiles of both $\Sigma$ and $\mathcal{P}$ by the function $f(r)$, where:
\begin{equation}
    f(r)\! =\! 10^{-4} + (1\!-\!10^{-4})\exp{ \lbrace -(2a/\sqrt{r^2+r_\mathrm{soft}^2})^{30}\rbrace}.
\end{equation}

The gravitational field of each individual BH is modelled by a Plummer potential,
\begin{equation}
    \Phi_n = -\frac{GM_n}{\sqrt{r_n^2+r_{\mathrm{soft}}^2}},
\end{equation}
where $M_n$ is the mass of the $n$th BH and $r_n$ is the distance from a field point to the $n$th BH.

We use a Cartesian grid, with a square domain of side length $20\,a$. We first run our simulations starting at low resolution and then, as the simulation settles into a steady state, we uniformly and globally refine the grid multiple times to increasingly higher resolution. In practice, our first resolution is $\Delta x\equiv0.04\,a$ and we run for roughly 5 viscous times, equivalent to 3000 orbits, where a viscous time is defined by $t_{\rm \nu_\mathrm{visc}} = 2/3\ r^2 / \nu_\mathrm{visc}$ at $r=a$. We refine the grid uniformly to a resolution of $\Delta x =0.02\,a$, keeping the sink size fixed. We continue this process, reducing the run-time as we progress, until we reach a final resolution of $\Delta x=0.005\,a$. Due to the very high computational expense at this final resolution, we begin analysis after one tenth of a viscous time has passed, and use the subsequent 8 orbits, sampled 1000 times per orbit, as input for all our \texttt{RAPTOR} models.

\subsection{Adaptive general-relativistic ray tracing setup}
\label{subsec:RaptorSetup}

In this section we discuss the general-relativistic ray tracing code \texttt{RAPTOR}, which consists of a grid-based camera constructed out of pixels. As the code integrates the geodesic equation, it simultaneously solves the radiative transfer equation backwards in time to find the amount of radiation per pixel at a given frequency. To accomplish this, a plasma model, spacetime metric, and radiation model all are needed. How they are determined is discussed below.

All values for the BHs and gas are taken directly from \texttt{Sailfish} and converted to a format understood by \texttt{RAPTOR}. The volumetric mass density $\rho(x,y,z)$ is approximated by invoking isothermal hydrostatic equilibrium in the vertical direction,
\begin{equation}
    \rho(x,y,z) = \rho_0(x, y) \, \mathrm{exp}(-z^2/(2h(x,y))^2),
\end{equation}
where the disk's half-thickness $h$ is given by Eqn.~\ref{eq:disk-height}, and $\rho_0(x,y)$ is determined by requiring $\Sigma(x, y) = \int_{-\infty}^{\infty} \rho(x,y,z) \,dz$. The velocity is taken to be vertically constant. The local effective temperature $T_{\rm eff}$ is determined from the local hydrodynamic fields $\mathcal{P}$ and $\Sigma$ via Eqn. \ref{eq:cooling}. Lastly, following the procedure developed in \cite{westernacher-schneider2022} for approximating the presence of radiation pressure in post-processing \citep[see also][]{WS+2023, Krauth2023}, we also scale down $\rho$ and $T_{\rm eff}$ such that the resulting accretion rate is the aforementioned $\dot{M}=0.19\,\dot{M}_{\rm Edd}$ \citep[see equations 14, 24, and 25 in reference][]{westernacher-schneider2022}.

To construct an approximate binary metric, we superpose two Cartesian-Schwarzschild metrics \cite{davelaar2022} (i.e.~Cartesian Kerr-Schild metrics with zero spin). The covariant superposed metric, $g_{\mu\nu}$, in units $G=M=c=1$, is given by
\begin{equation}
g_{\mu\nu} = \eta_{\mu\nu} + h^p_{\mu\nu} + h^s_{\mu\nu},
\end{equation}
where subscripts $p$ and $s$ indicate the primary and secondary BHs respectively. The metric is a superposition of the Minkowski metric, ${\eta_{\mu\nu}={(-1,1,1,1)}}$, and two curvature terms $h^{p/s}_{\mu\nu}=f^{p/s} l_\mu^{p/s} l_\nu^{p/s}$ for the BHs, where $f$ is a scaling factor, and $l_\nu$ is a null covector. The source terms are shifted with respect to the original spatial coordinates $\vec{x}$ via $\vec{x}_{p/s} =  \vec{x} - \vec{x}^{bh}$ with $\vec{x}^{bh}_{p/s}$ the position vector of the BH.
The scaling factor is $f = 2 / r$, and the null covector $l_\nu$ with zero spin is given by

\begin{eqnarray}
l_\nu &=  \begin{pmatrix}
1\\
x/r \\ y/r \\ z/r
\end{pmatrix}
\end{eqnarray}
where $r$ is the radial coordinate, equivalent to the radius in spherical Schwarzschild coordinates. The factors $f$ and $l_\nu$ always use the variables corresponding to their respective primary or secondary BH. The contravariant metric is similarly defined as
\begin{equation}
g^{\mu\nu} = \eta^{\mu\nu} - h^{\mu\nu,p} - h^{\mu\nu,s},
\end{equation}
with $h^{\mu\nu,p/s}=f^{p/s} l^{\mu,p/s} l^{\nu,p/s}$. Here $l^{\nu,p/s}$ is identical to $l_\nu^{p/s}$ except that the temporal component changes sign. In order to compute Doppler shifts, we need a four velocity $u^\mu$, since $\nu_\mathrm{p} = -k_{\mu} u^{\mu}$, where $k_{\mu}$ is the wave 4-vector and $\nu_\mathrm{p}$ is the photon frequency in the plasma frame. While the spatial components of the four-velocity are entirely determined by the hydrodynamic simulation, we need to compute the time component which we do through the $u^\mu u_\mu = -1$ normalization.

As the code integrates the geodesic equation backwards in time, from the camera to the simulated gas, it simultaneously computes the emission received at every pixel for multiple frequencies. We assume that the emission is purely thermal black body emission emitted at a photosphere at which the optical depth exceeds $\tau_\mathrm{eff_\nu}>1$. The optical depth is computed along the ray by $\tau_\mathrm{eff_\nu}=\int^{\lambda^{\rm obs}}_\lambda \nu \alpha d\lambda$. Here $\alpha$ is the absorption coefficient. For the gas to be considered thermalized, $\tau_\mathrm{eff} = \sqrt{\tau_\mathrm{a}(\tau_\mathrm{a}+\tau_\mathrm{s})}~\gg~1$, where $\tau_\mathrm{a}$ and $\tau_\mathrm{s}$ are the absorption and scattering optical depth, respectively. In practice, we find the effective optical depths exceed unity throughout our simulations. Beyond that however, due to electron scattering being the dominant source of opacity, we can safely disregard absorptive mechanisms, as we do in our analysis. Therefore we use $\alpha_\nu=\kappa\rho$, where $\kappa$ is the Thompson opacity, and $\rho$ is the gas density. We use the "fast-light" approximation, in which the metric and plasma are assumed static during ray-tracing. The integral goes from the "camera" to the source along the geodesic, and terminates as soon as $\tau_\mathrm{eff_\nu}\geq 1$, where we set the intensity to a black body using the local effective temperature $T_{\rm eff}$ as described above. No other emission is accounted for between the photosphere and the camera.

The ray-tracing calculation includes emission from gas within a square region, centered on the binary, with a side-length of $3\,a$. This region is large enough to properly synthesize images of the minidisks, which are the main regions of interest, being the light sources to be lensed by the foreground BH. We employ a block based adaptive mesh refinement (AMR) algorithm for the camera grid. The camera is split in blocks of equal pixel size and number, after which the code computes the emission in a given block at the lowest resolution. The AMR algorithm then refines this block (increasing the resolution by a factor and splitting the block twice in both directions), if the gradient of the intensity exceeds a predefined threshold ($\Delta I/I = 10^{-3}$). For a summary of the AMR procedure and a resolution convergence test, see Appendix~\ref{app-a}. A full description is outlined in detail in Paper~I.

Light curves are subsequently created by integrating the images and aggregating the total flux at multiple frequencies: $0.54\,\mu$m (optical), 83 eV (hard UV), 0.3 keV (soft X-ray), and 1.0 keV (harder X-ray). Once computed, the light curves are phase-folded on the timescales of half an orbital period for the total number of orbits for each model listed in Table~\ref{tab:Models}. We also compute the amplification of the SLF compared to the baseline mean luminosity, which is calculated by normalizing the entire phase-folded light curve by the average luminosity for all times outside of the peak of the SLF. Because the start and end of the flare is somewhat arbitrary, this amplification should be regarded as a rough estimate.

\subsection{Models}
\label{subsec:Models}

We use \texttt{Sailfish} to simulate accretion onto an equal-mass binary black hole on a circular orbit with total mass $M_{\rm bin}=10^6$~\(\rm M_\odot\), near the peak of LISA's anticipated frequency range \citep{amaro-seoane2017}. The binary has a separation of $50\,r_{\mathrm g}$. The thermodynamic conditions in the disk are selected to achieve a characteristic orbital Mach number $\mathcal{M} \sim 30$ (see Eqn. \ref{eq:cooling}). The high-cadence simulation snapshots span a total of 8 binary orbits. As a single simulation can be used to calculate the SLF that would be seen by observers at any viewing angle, we analyze a differing number of these orbits for several distinct \texttt{RAPTOR} configurations. In an equal-mass binary, the CBD is known to develop a significantly eccentric morphology in the binary's vicinity. The eccentric CBD adds a second viewing angle parameter, namely the disk's argument of periapse. We have carried out all of our ray-tracing simulations with the line-of-sight oriented perpendicular to the eccentricity vector of the CBD, with the disk periapse towards the right side of our synthesized images. As the equatorial plane of our disk lies in the $x-y$ plane, this corresponds to our camera's position lying in the $y-z$ plane. 

Our fiducial ray-tracing calculation places the camera at a line-of-sight $\theta = 86^{\circ}$ away from the orbital angular momentum vector. We have found this to be the nearest to edge-on viewing that is possible before the lines-of-sight to the minidisks become obstructed by the (optically thick) CBD.

SLF features are generally most prominent for nearly edge-on observers. However if $\theta$ is too close to $90^\circ$, the physical thickness of the CBD can obscure the view of the bright minidisks. Thus a thick CBD could obstruct the view of what would otherwise be the most dramatic possible SLF's. But realistic disks may be warped \citep[e.g.][]{Nixon2016}, and are likely thinner geometrically than those we could feasibly simulate in this study. To examine SLF's produced by systems with a very thin CBD, we computed models with the disk height $h$ artificially reduced by a factor of 3, while holding constant other hydrodynamic parameters including $\Sigma$ (note that the thinner disk implies a higher orbital Mach number than was used in the \texttt{Sailfish} calculation).

To perform a ray-tracing calculation on an artificially warped disk, we apply a local spatial rotation of the position and velocity of the gas parcels, in the direction around the $x$-axis with a rotation matrix $\vec{X'}=R(\theta_\mathrm{w})\vec{X}$. Here $\theta_\mathrm{w} = f(r)\theta_\mathrm{max}$ is the warping angle at each point and $\theta_\mathrm{max}$ is the maximum warping angle. The smoothed step function $f(r) = 0.5\{1+\tanh [(r-r_0)/\Delta R]\}$ produces the warp outside of radius $r_0=58.3\,r_\mathrm{g}$. The adopted smoothing size is $\Delta R = 16.6\,r_\mathrm{g}$. The disk warping is not dynamically self-consistent, but gives a reasonable approximation of the anticipated effects. The parameters used to artificially warp the disk are in line with predictions \cite{Rabago2023}. 

To lower the computational expense, we limit the number of orbits we phase-fold and analyze in \texttt{RAPTOR} once the SLF morphology has stopped changing appreciably (unless specified otherwise; see the discussion around Figure~\ref{fig:fid_scale0p3_warp_compare}). Or, in the case of the lower inclination angle, we run at a lower temporal resolution, as the primary purpose for the higher cadence is to search for the dipping feature in the signal, caused by lensing of the background BH shadow, which is not possible at such a low angle [e.g. Paper I]. This information and other details for all models are summarized in Table~\ref{tab:Models}.
    
\begin{table*}
\caption{Our suite of simulations, with the parameters they depend on. $\theta_\mathrm{inc}$ represents the viewing angle for the observer. $h/h_\mathrm{fid}$ is the reduction factor of the disk scale height. $\theta_\mathrm{max}$ is the maximum angle (in degrees) by which the CBD is warped. Snapshots/orbit is the amount of checkpoints per orbit that was analyzed within \texttt{RAPTOR} and $N_\mathrm{orbits}$ is the number of total orbits analyzed within \texttt{RAPTOR}. Boldface text denotes which parameters have been modified from the fiducial model, shown in the first row.}
\centering
\setlength{\tabcolsep}{10pt}
\begin{tabular}{l| l| c c c c c}
 \hline \hline
 Model &Description &$\theta_\mathrm{inc}$(deg) &$h/h_\mathrm{fid}$ &$\theta_\mathrm{max}$(deg) &Snapshots/orbit &$N_\mathrm{orbits}$\\
 \hline
 Fiducial &Fiducial Model &$86^{\circ}$ &1 &$0^{\circ}$ &1000 &8\\
 I80 &Lower viewing angle &\boldmath{$80^{\circ}$} &1 &$0^{\circ}$ &\boldmath{$200$} &\boldmath{$6$}\\
 h1/3\_I86 &Scaled $h$ by 1/3 &$86^{\circ}$ &\boldmath{$1/3$} &$0^{\circ}$ &1000 &\boldmath{$2$}\\
 h1/3\_I89 &Scaled $h$ by 1/3, $89^{\circ}\ \theta_\mathrm{inc}$ &\boldmath{$89^{\circ}$} &\boldmath{$1/3$} &$0^{\circ}$ &1000 &\boldmath{$4$}\\
 w4\_I86 &Disk warped by $4^{\circ}$ &$86^{\circ}$ &1 &\boldmath{$4^{\circ}$} &1000 &\boldmath{$2$}\\
 w4\_I90 &Disk warped by $4^{\circ}$, $90^{\circ}\ \theta_\mathrm{inc}$ &\boldmath{$90^{\circ}$} &1 &\boldmath{$4^{\circ}$} &1000 &\boldmath{$4$}\\
 \hline \hline
\end{tabular}
\label{tab:Models}
\end{table*}

\section{Results and discussion}
\label{sec:Results}

In the following sections we present our results first for our fiducial model and then for its several variations. We discuss the implications, similarities, and differences throughout. Although we expect that the variability of the $0.54\,\mu$m optical emission is well-captured in our results, its average level is not, because we are not capturing a significant amount of low-variability $0.54\,\mu$m emission coming from the CBD at radii beyond our computational domain. In what follows, this caveat should be kept in mind when any comparisons are made to the average level of $0.54\,\mu$m emission. Emission at higher photon energies is captured more completely, since it is dominated by regions within our computational domain.

\subsection{Fiducial model}
\label{subsubsec:Fid}

\begin{figure*}
    \centering
    \includegraphics[width=0.95\textwidth]{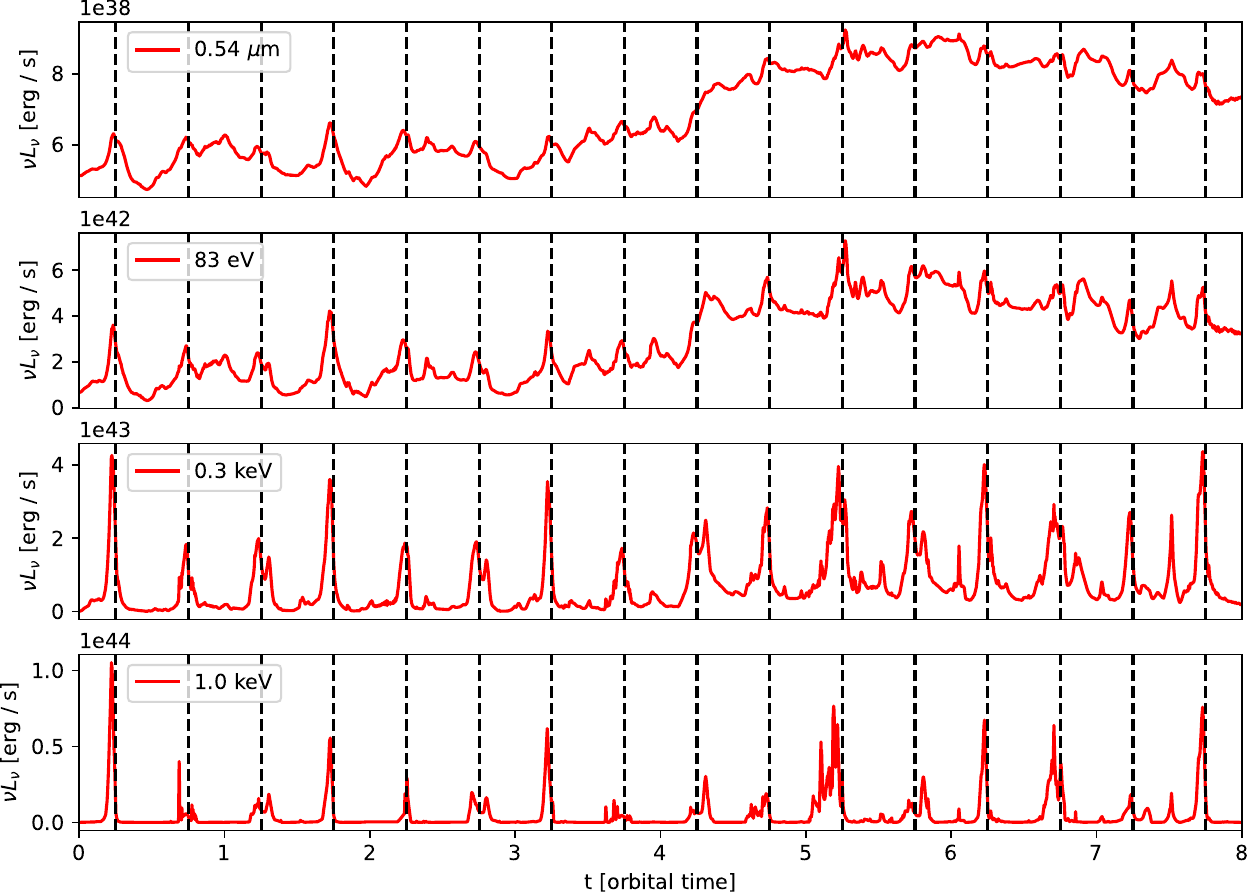}
    \caption{Full fiducial 8 orbit non-phased-folded luminosity for the four frequencies studied here as listed. The black dotted lines correspond to when the BHs are aligned with the observer's line-of-sight.}
    \label{fig:full8}
\end{figure*}

Although much of this study will use phase-folded light curves to extract an SLF signature, we find it instructive to first start by showing a non-phase-folded light curve from our fiducial model in Figure~\ref{fig:full8}. The four frequencies studied are shown here, with the black dotted lines corresponding to when the BHs are aligned with the observer's line-of-sight. As time progresses, the baseline flux at lower photon energy rises somewhat, due to the bright lump orbiting into view along the cavity wall. We note that the higher frequencies, the 0.3 keV and 1.0 keV produce the brightest emission, and spike significantly when we would expect an SLF. The emission at 0.54\,$\mu$m and 83 eV also often spikes coincident with this alignment, but has more overall noise. We will however momentarily phase-fold these frequencies, in which the behavior for all four will become more clear. We also see that the spikes in 1.0 keV emission are more dramatic than the 0.3 keV emission, and the spike amplitudes are almost quasi-periodic. This quasi-periodicity arises because the brightest flares occur when the gas whose emission is being lensed is at its hottest, like when mass-trading or when higher accretion on to the background minidisk has recently occurred; this enhanced minidisk brightness is not always coincident with the SLF. 

\begin{figure}
    \centering
    \includegraphics[width=0.47\textwidth]{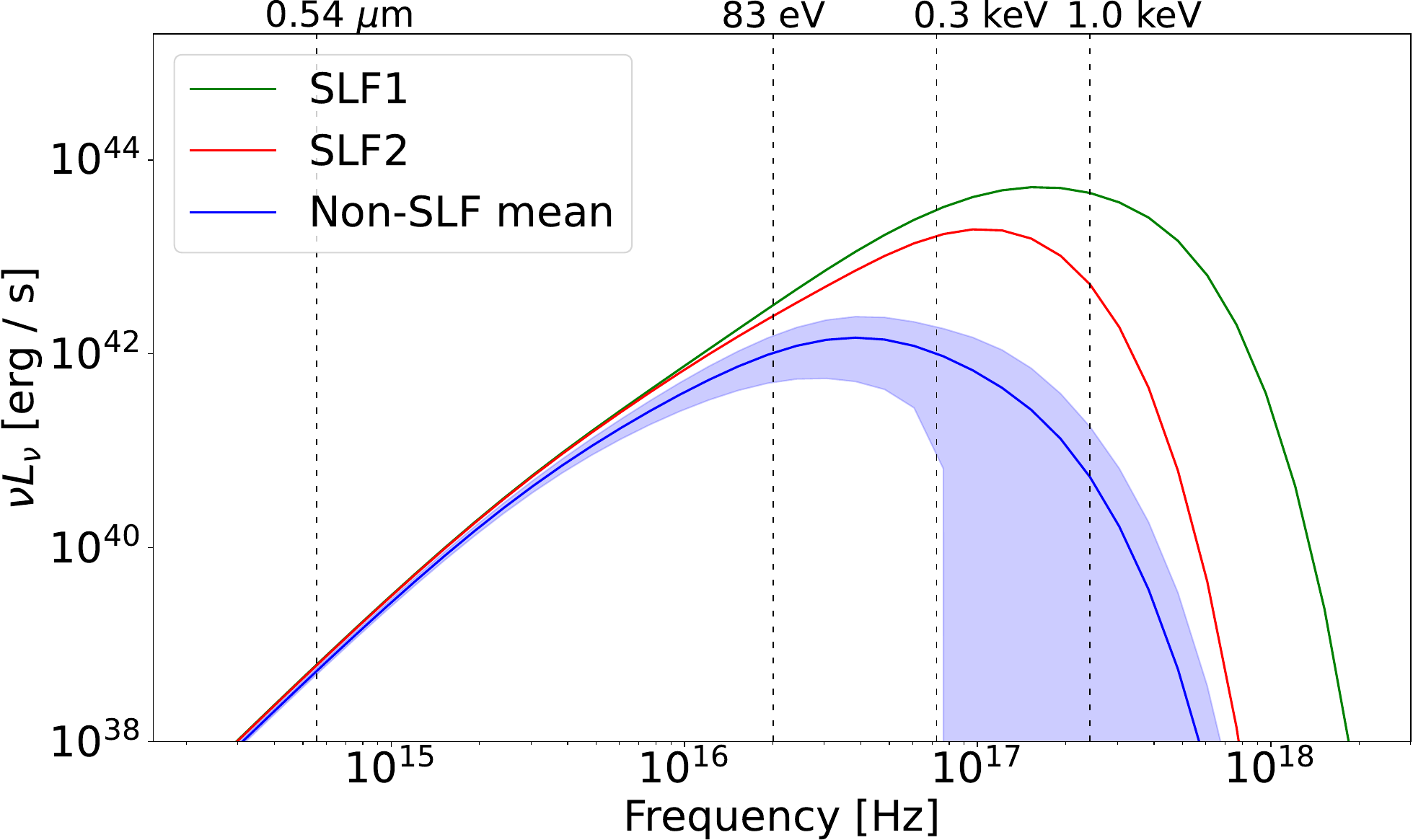}
    \caption{Spectra in our fiducial model computed during an orbit's two SLFs (green and red) and the average spectrum with its standard deviation (blue with a shaded blue region) for all times outside of the SLF peak. The spectra during the SLFs are both more luminous and harder, several $\sigma$ above the non-SLF mean, and are thereby likely to be easily distinguishable.}
    \label{fig:fid_spectra}
\end{figure}

\begin{figure*}
    \centering
    \includegraphics[trim={2.6cm 0 0 1.5cm},clip,width=0.95\textwidth]{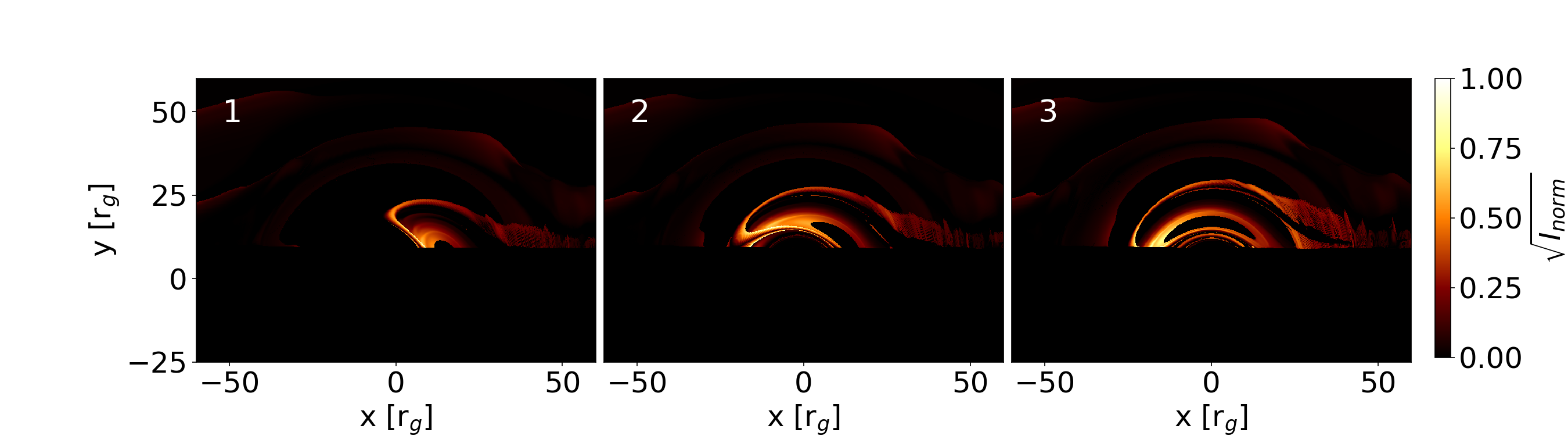}
    \caption{Snapshots of the UV surface brightness distributions (at 83 eV) in our fiducial model at three different times. The color scale is the square root of the normalized intensity for contrast. These times correspond to those during the phase-folded light curve marked in the upper right panel of Fig.~\ref{fig:fid_phasefold}. We see significant blockage of the minidisk emission by the CBD, yet the lensing is able to bend the emission over this blockage.}
    \label{fig:123}
\end{figure*}

Figure~\ref{fig:fid_spectra} shows the spectra from our fiducial model. Because the spectrum is highly time-dependent, we have isolated three important times during an orbit. The blue curve, with a shaded blue region, represents the average spectrum and its standard deviation, for all times during an orbit which are not during an SLF. Conversely, the green and red curves represent the spectra for the first and second flares in an orbit, respectively. We note that during the flare, the spectrum is not only harder in X-rays, but significantly more luminous, several $\sigma$ above the non-SLF average. The vertical dotted lines represent the photon frequencies we analyze in the rest of our study.

\begin{figure}
    \centering
    \includegraphics[width=0.47\textwidth]{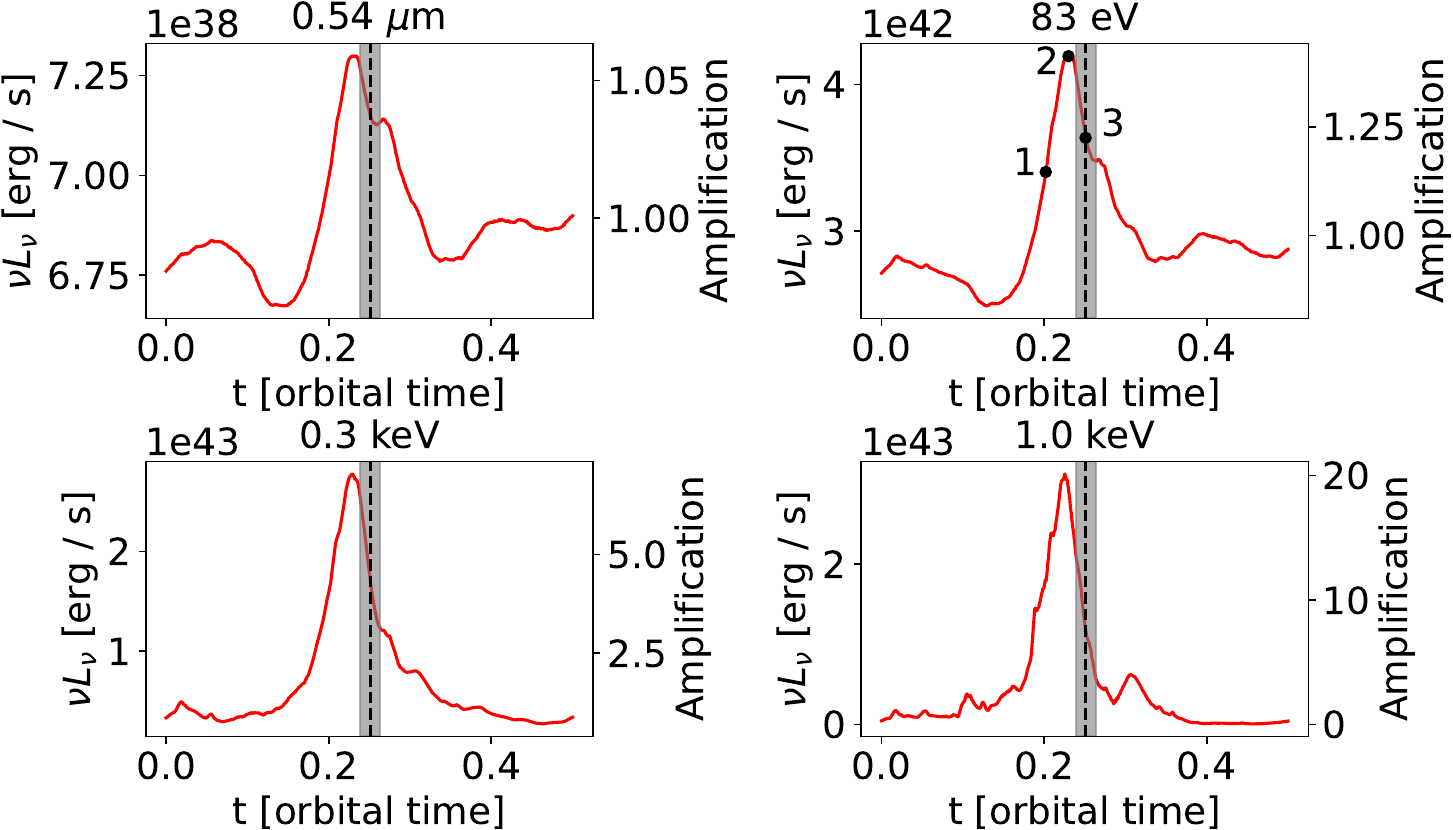}
    \caption{Light curves in the fiducial model, phase-folded over 8 orbits. The vertical dotted black lines mark the moments of direct alignment to the line-of-sight and the gray regions around them represent the expected transit duration of an axisymmetric shadow. We see increasing overall amplification but a decreasing depth of the "dip" with energy. The moments labeled 1, 2, 3 in the upper right panel correspond to the images in Fig.~\ref{fig:123}.}
    \label{fig:fid_phasefold}
\end{figure}

Figure~\ref{fig:123} shows zoomed snapshots of the surface brightness in our fiducial model at the hard-UV energy of 83 eV at three different times. The color scale represents the square root of the normalized intensity (the intensity in each pixel normalized by the brightest pixel) for contrast. These times correspond to those marked in the upper right panel of Figure~\ref{fig:fid_phasefold}, with "1" corresponding to the rise of the SLF, "2" corresponding to the left-side peak in the SLF, and "3" corresponding to the dip feature in the center of the SLF. The vertical dotted black lines in Figure~\ref{fig:fid_phasefold} represent the moments of perfect alignment to our line-of-sight, and the gray regions around them represent the duration of the BH shadow transit, whose orbital phase width is given by $\Delta\phi=C/(2\pi a)$ where a is the binary's orbital separation and $C=\sqrt{r_\mathrm{shadow}^2-d^2}$ is the chord length of the shadow at the inclination angle of the system, where $r_\mathrm{shadow}=2.6\,r_\mathrm{s}$, and $d=a\tanh({\pi/2-\theta})$ is the vertical distance from the center of the shadow to the focal point of the lens. These figures allow a direct comparison with the semi-analytical model of Fig.~1 in Paper~II. 

The most clear difference between Fig.~1 in Paper~II and our Figure~\ref{fig:123} is the significant obscuration of the emission from the inner regions of the system, appearing as the entirely dark lower halves of the images in Figure~\ref{fig:123}. This is caused by the CBD blocking the line of sight, something not included in the semi-analytical model of Paper~II. During the SLF, emission from the background minidisk is lensed above the CBD. Despite this large difference with Paper~II, the similarities are also striking. E.g.~we see that the rise of the flare begins just as the minidisk of the background BH becomes lensed by the foreground BH. We also note that the peak in the light curve corresponds to lensing of the brightest, innermost region of the minidisk. Then, after we pass this brightest region, there is a dip in the flux, where the center coincides with the alignment of the BHs with our line of sight, indicating this is when the foreground BH is now fully lensing the shadow of the background BH. Our Figure~\ref{fig:123} also shows emission from a rainbow-shaped region above the minidisks, which is the shock-heated cavity wall of the CBD.

Note the asymmetry in the double peak of our curve in Figure~\ref{fig:fid_phasefold}, which arises due to Doppler effects. The left side of the disk, which is lensed first, is rotating towards the observer and is hence blue-shifted. By contrast, the right side of the disk, which is lensed last, is red-shifted. Because of the shape of the Wien tail for the black-body spectrum, this causes a brightening in the left side of the flare, and, to a smaller degree, a dimming in the right side of the flare. We see this effect increases with photon energy in Figure~\ref{fig:fid_phasefold}. This makes sense, since the higher-energy emission comes from a smaller inner region, where internal velocities in the minidisk regions are higher. This effect has been explored in Fig.~4 of Paper~I. One effect considered in Paper~I that we did not include here is "slow-light," where the finite propagation speed of light is take into account; we use the "fast-light" approximation in this work (effectively corresponding to an infinite speed of light). The inclusion of slow-light effects in Paper~I resulted in re-balancing of the brightening and dimming of the left versus right side of the SLF, which tends to help recover the structure of the dipping feature. We might therefore expect that, for our case, slow-light would produce similar re-balancing, which may make the SLF dip more observable.

Figure~\ref{fig:fid_phasefold} shows the SLFs are present in each band, but they are highly chromatic. While the optical, 0.54\,$\mu$m emission arguably shows the most distinguishable features of the flare and dip, the amplification is only $\sim5\%$, much lower than the other frequencies. This is because most of the optical emission originates from the CBD, which is not predominantly within the Einstein radius of the foreground BH. Hence, only the small fraction of the optical emission, coming from the background minidisk, can be lensed and contributes to the flare. Considering instead higher energy emission, the converse happens, since the fraction of the luminosity originating from the minidisk increases with frequency \cite{Krauth2023}. Because the minidisks are the most strongly lensed regions, this suggests lensing amplification should increase with frequency. The hard-UV frequency (83 eV) is indeed significantly brighter, showing a pronounced SLF flare and dip, and has an amplification of 50\% above the baseline flux. The 0.3 keV and 1.0 keV energies are the brightest and most strongly amplified, with SLF fluxes $\sim$7 and $\sim$20 times their respective baselines. However, as dipping features becomes less and less prominent, eventually the enhanced Doppler effects are steep enough that the dip is washed away. Even still, SLFs may be easiest to detect at these frequencies because of the extreme amplifications. 

\begin{figure}
    \centering
    \includegraphics[width=0.47\textwidth]{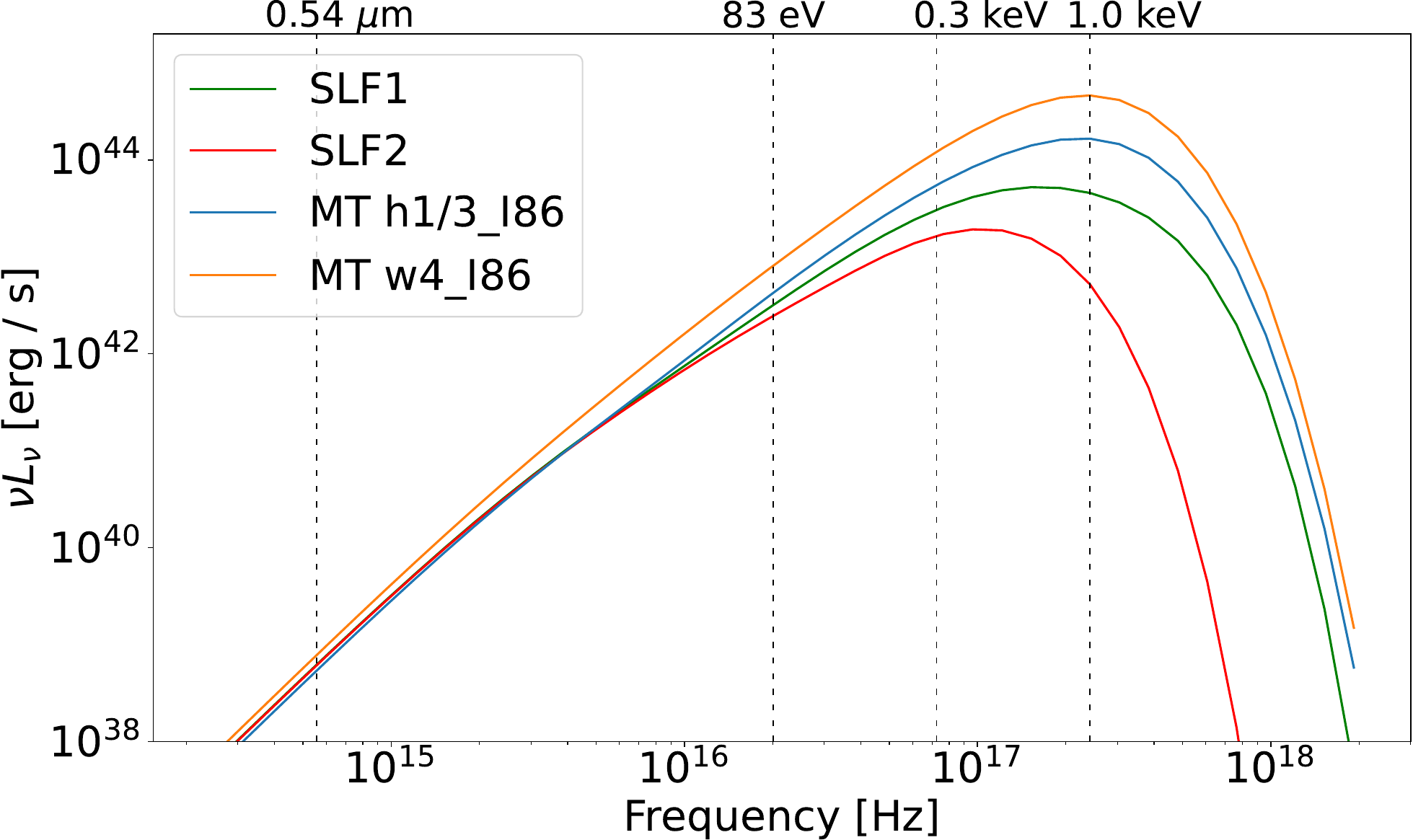}
    \caption{The same spectra of the SLFs in our fiducial model (green and red) as seen in Figure~\ref{fig:fid_spectra}, but now overlaid with the spectra during the mass-trading events (abbreviated as MT) in the models h1/3\_I86, and w4\_I86. The spectra of the mass-trading events are both harder and brighter than those of the SLFs.}
    \label{fig:slf_mt_spectra}
\end{figure}

\subsection{Model comparisons}
\label{subsec:ModCompare}

In the following sections, we examine the effects of thinning and warping the disk, as well as different viewing inclination angles, comparing them to our fiducial model and each other.

\subsubsection{Effects of disk thinning and warping}
\label{subsubsec:ThinWarp}

\begin{figure*}
    \centering
    \includegraphics[trim={2.6cm 0 0 1.5cm},clip,width=0.95\textwidth]{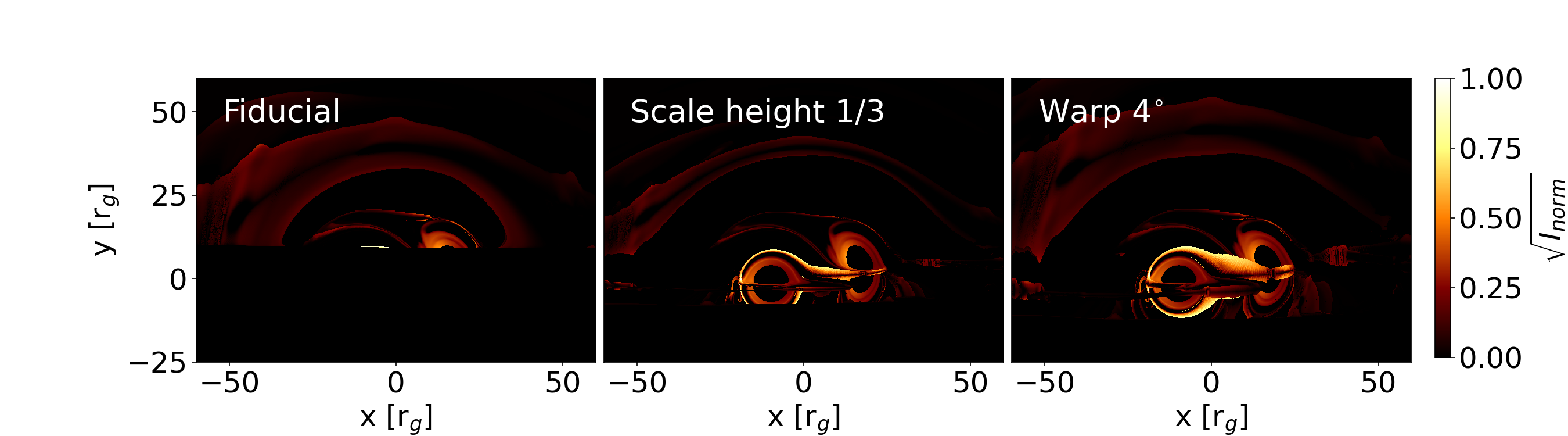}
    \caption{UV images in our fiducial (left), thinned disk (h1/3\_I86, center), and warped disk (w4\_I86, right) models at 83 eV at the same mass-trading moment. The color scale is the square root of the normalized intensity for contrast. The fiducial model blocks most of the minidisk emission, whereas the thinned disk model reveals significantly more, and the warped model has the clearest view.}
    \label{fig:mass_trade_compare}
\end{figure*}

Thinner or warped disks can make SLFs visible from more edge-on viewing angles, which would otherwise be obscured by the CBD. However, before we examine more edge-on configurations, we first examine what the effects of thinning and warping are while keeping the viewing angle constant.

One of the brightest moments during an orbit is when the minidisks trade mass \cite{westernacher-schneider2022}. In Figure~\ref{fig:slf_mt_spectra} we plot spectra for the two SLFs previously shown in Figure~\ref{fig:fid_spectra}, and compare them to the spectra during a mass-trading event in the two alternative models  h1/3\_I86 and w4\_I86, corresponding to disk thinning and warping, respectively. Much more of the region near the BHs is made visible by the thinning and warping, as seen in the three zoomed snapshots of Figure~\ref{fig:mass_trade_compare}, all shown at the moment of a mass-trading flare. An obvious difference between these models is the fraction of the emission blocked by the CBD. In the fiducial model, emission from the mass-trading event is almost entirely obscured (left panel). With the thinned disk, we see most of the radiation from the event, with only a small portion of the lower half of the foreground BH's minidisk occluded (center panel). In the warped model (right panel), we see nearly the full event.

\begin{figure}
    \centering
    \includegraphics[width=0.47\textwidth]{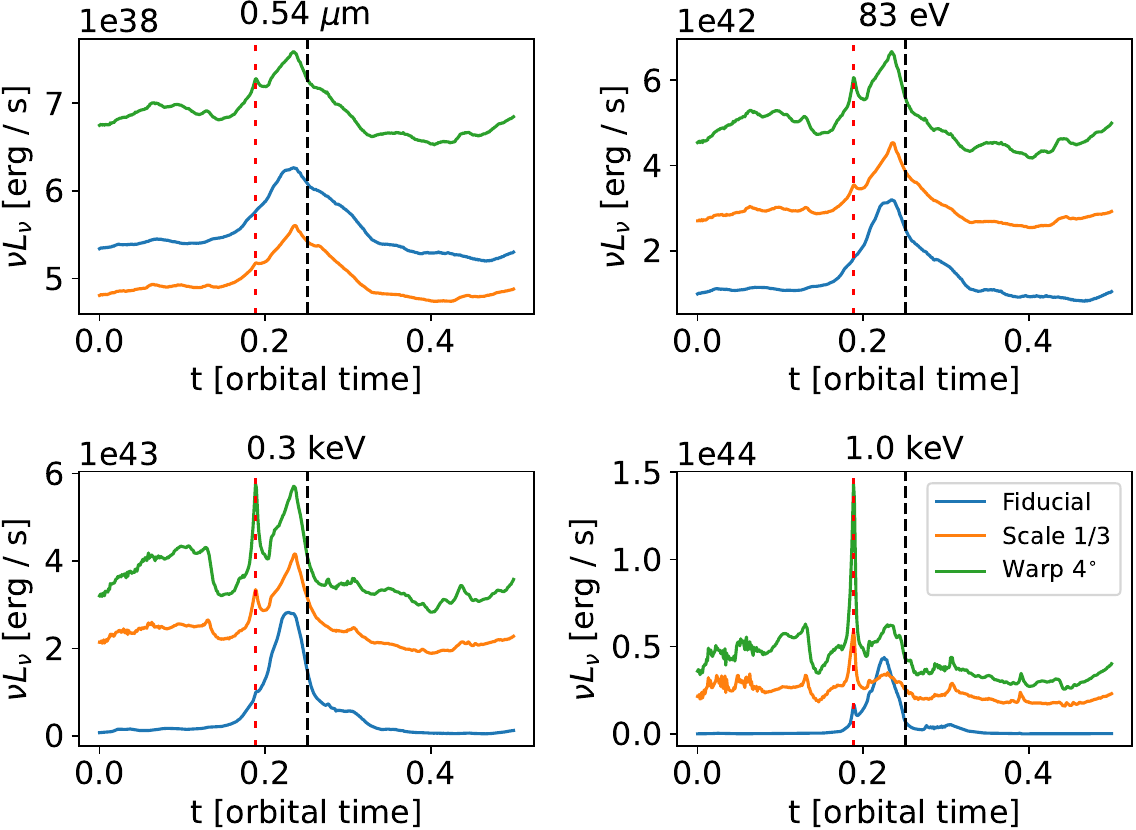}
    \caption{Light curves in the fiducial model (blue), the thinned disk model (h1/3\_I86, orange), and the warped disk model (w4\_I86, green), phase-folded over 2 orbits. While each model retains a SLF, the thinned and warped models also show significant lensing amplification of the mass-trading event. The red [black] vertical dashed lines mark the time of mass-trading [alignment of the BHs to our line-of-sight].}
    \label{fig:fid_scale0p3_warp_compare}
\end{figure}

Figure~\ref{fig:slf_mt_spectra} shows that when the mass-trading event is visible, its spectrum is both brighter and harder than the SLFs. Figure~\ref{fig:fid_scale0p3_warp_compare} shows light curves phase-folded over two orbits, in which a significantly bright mass-trading event happens to occur during the first orbit. Note we only phase-fold over 2 orbits here due to computational expense, so these light curves would resolve the structure of the SLF if phase-folded further. But phase-folding over two orbits is enough to make some interesting comparisons. The vertical black dotted lines mark the alignment of the BHs with the line of sight, whereas the red dotted lines mark the moment of mass trading. Even though mass-trading is not periodic on the orbital time-scale, we can still see its presence after phase-folding over two orbits, and we see how this manifests in the different models. 

In the fiducial model (blue), there is a slight additional feature present in the light curve at the moment of mass-trading at 1.0 keV. In the thinner disk model, however, a spike is present, becoming increasingly prominent at higher photon energies. The warped model shows the most significant mass-trading spike, dwarfing the SLF peak at the highest energy. The weaker mass-trading spike in the thinner disk model compared to that of the warped model is likely due to two reasons. First, the warped model simply has a more complete view of the binary; second, rays coming in at oblique angles with respect to the binary plane are more likely to miss a thinner disk near its edges, i.e. a thinner disk's projected area on the camera is reduced.

Continuing with Figure~\ref{fig:fid_scale0p3_warp_compare}, a peculiar and interesting feature presents at 1.0 keV. We notice that during the SLF, the thinner disk model has a smaller peak than the fiducial model. Naively one might expect the thinner disk model would have a greater SLF peak because more of the background minidisk is visible. But similarly to the mass-trading flare discussed above, there is less emission to be lensed because reducing the disk scale height leads to a reduced projected surface area of the minidisk. This implies the counterintuitive fact that a higher Mach, thinner disk does not necessarily lead to a brighter SLF peak. While the thinner CBD helps to observe the inner emission, a thinner minidisk provides less emission to be amplified during an SLF, even when it can be viewed. Perhaps a favorable situation for observing an SLF in reality would be when the CBD is thin, but the minidisk (that has been heated by additional shocks) is not. Although, as obvious in the 1.0 keV light curve in the fiducial model, a large contrast between the baseline flux and SLF peak is most favorable for SLF detection. In the fiducial model, this contrast exists because the CBD lowers the baseline flux by blocking the non-SLF minidisk emission, but does not block the SLF emission.

The mass trading event is almost entirely obscured in the fiducial model shown here. But it is in principle possible for the mass trade and the SLF to coincide, since trading occurs at about $\sim 1.2$ times the orbital frequency in this model \citep[e.g.][]{westernacher-schneider2022, WS+2023}. While such a coincidence did not happen naturally in our models, we did test this by shifting the azimuthal angle of our camera; because the mass-trading flux is orders of magnitude larger than the SLF and does not necessarily become directly lensed itself, we found the spectrum to be comparable to that of a normal mass-trading flare. More importantly, since the mass-trade can occur at any moment during the SLF, it could contaminate a particular SLF signature. Phase-folding over sufficiently many binary orbital periods should reduce the complicating effect of mass-trading flares, as long such flares repeat with a period that differs sufficiently from the orbital and SLF periods.

\subsubsection{Thinned disk}
\label{subsubsec:Scale0p3}

\begin{figure}
    \centering
    \includegraphics[width=0.47\textwidth]{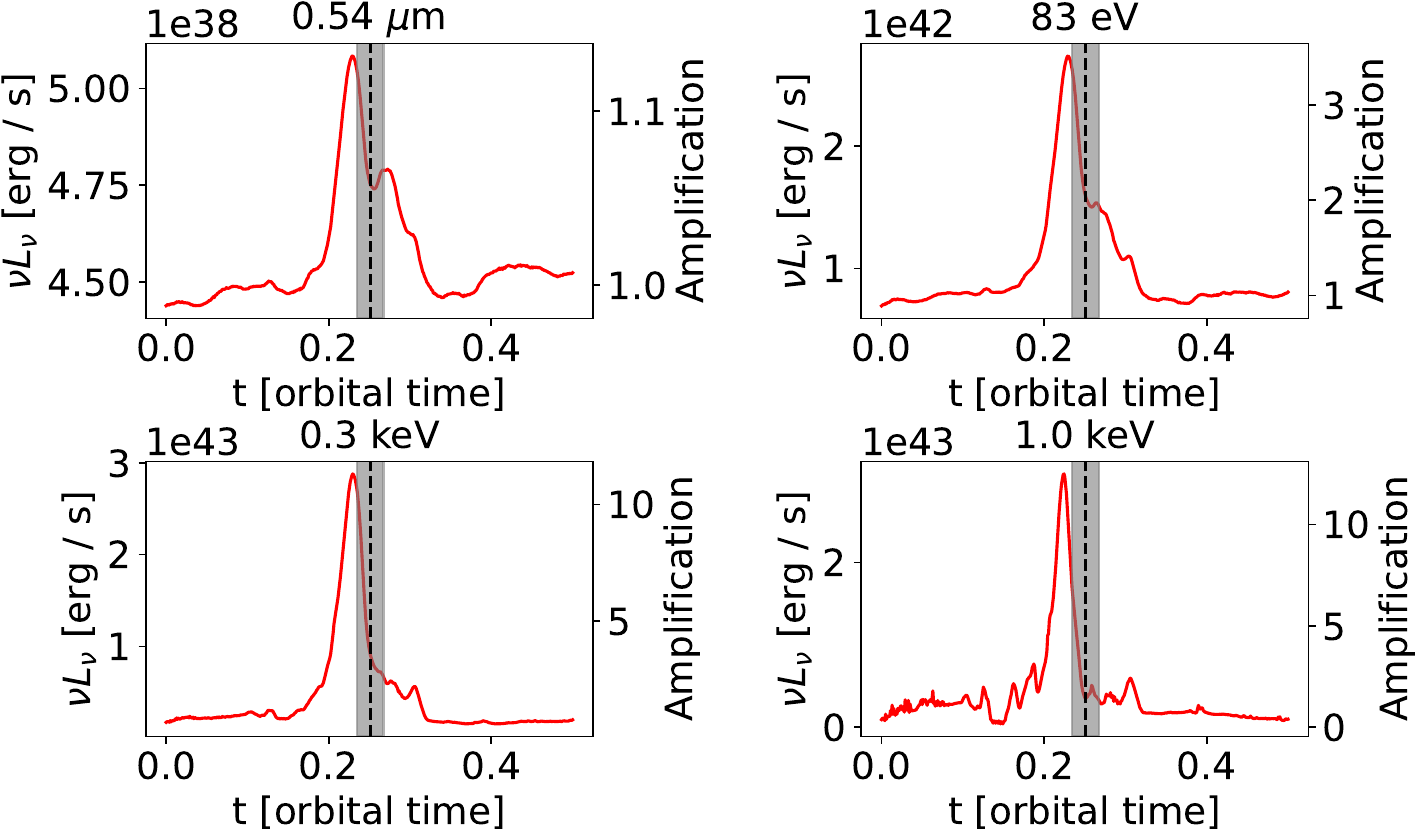}
    \caption{Light curves in the thinned disk model h1/3\_I89, phase-folded over 4 orbits. The dotted vertical black lines mark the moment of direct alignment of the two BHs with the line-of-sight and the gray region around them represent the transit duration of the BH shadow. We see increasing amplification and flare asymmetry, and less discernible "dips" as the photon energy is increased.}
    \label{fig:189_phasefold}
\end{figure}

Figure~\ref{fig:189_phasefold} shows the phase-folded light curve for model h1/3\_I89, in which the scale height is once again reduced by a factor of 3, but now the inclination angle has been increased to 89$^{\circ}$ (i.e.~more edge-on). With the higher inclination angle, the CBD again blocks a large portion of the emission from the minidisks outside the duration of the SLFs, namely the emission during mass-trading. As such, we recover a clean signature of the SLF, although with some differences. Perhaps the most notable difference is the amplification, which is nearly double that in our fiducial model for most energy values. This is likely because the increased inclination angle leads to the lens passing more centrally over the background minidisk, thereby amplifying its emission even further. Another, perhaps more nuanced difference, is the shape of the double peak. We now see a larger asymmetry between the left and right side peaks than in our fiducial model. Given that this difference comes from Doppler effects, when we have a higher inclination angle, the faster inner parts of the disks get a higher lensing weight, thereby increasing the Doppler effects. This unfortunately means that the dipping feature appears to wash out slightly sooner. Although harder to discern, the shadow size is also slightly wider than in the fiducial model, because the BHs are now more aligned to the observer. This means the lens passes over the shadow of the background BH more centrally, leading to a wider dip during the SLF. 

\subsubsection{Warped disk}
\label{subsubsec:Warped}

\begin{figure}
    \centering
    \includegraphics[width=0.47\textwidth]{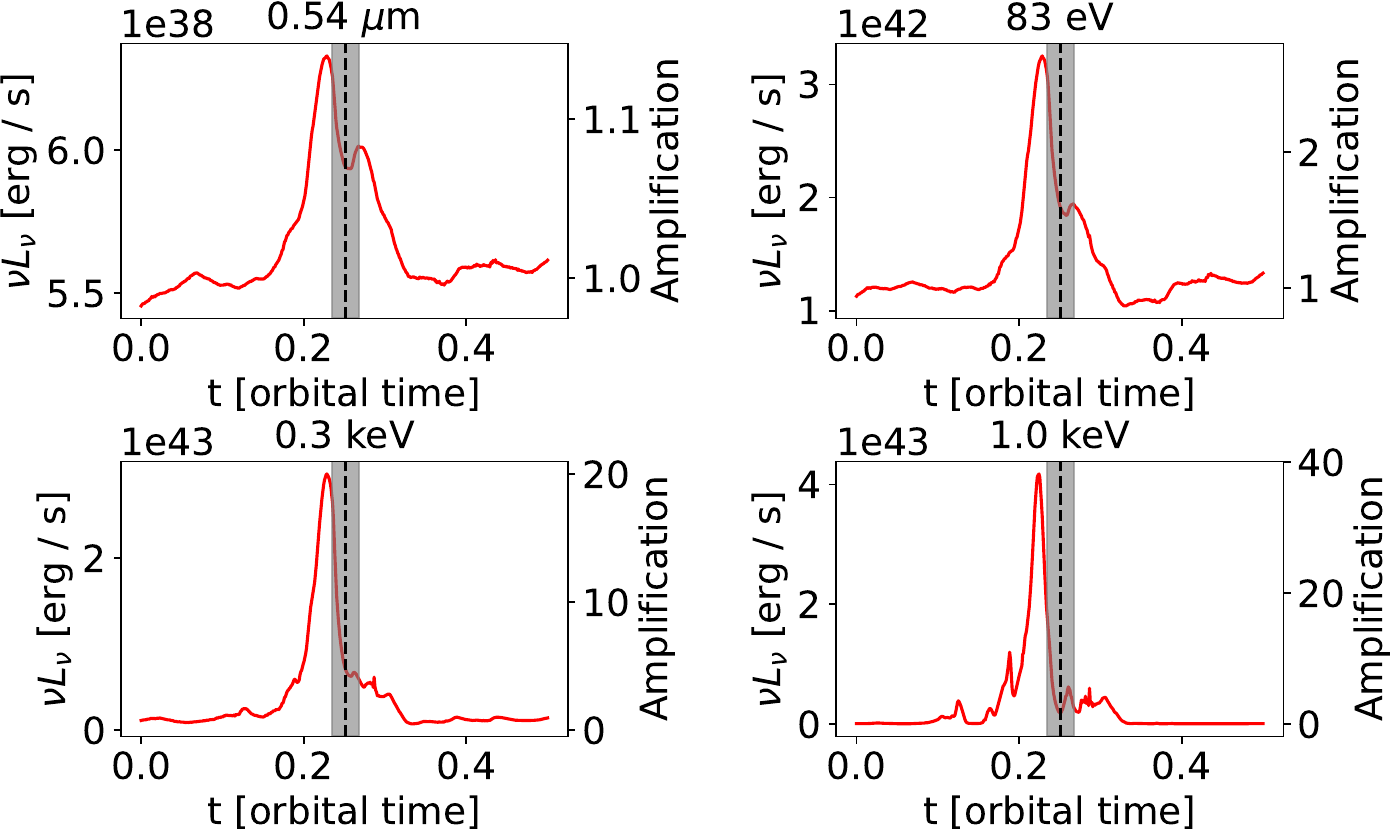}
    \caption{Light curves for the warped disk model w4\_I90, phase-folded over 4 orbits. The dotted black lines mark the moment of direct alignment of the two BHs with the line-of-sight and the gray region around them represent the expected transit duration of the BH shadow. We see increasing amplification and flare asymmetry, and less discernible "dips" as the photon energy is increased, similar to Fig.~\ref{fig:189_phasefold}, but with even higher amplification due to a decreased non-SLF mean.}
    \label{fig:904_phasefold}
\end{figure}

Figure~\ref{fig:904_phasefold} shows the phase-folded light curve for model w4\_I90, in which the disk is warped by 4$^{\circ}$, but now the inclination angle has been increased to 90$^{\circ}$ (edge-on). Again, the CBD blocks much of the emission not occurring during the SLF. The light curves are comparable to those in the h1/3\_I89 model, with a slight overall increase in the overall brightness. There is also a sizeable increase in the amplification at 1 keV. However, comparing the bottom right panels of Figures~\ref{fig:189_phasefold} and \ref{fig:904_phasefold}, we see that the peak flux is actually comparable; the main difference is that the warped disk has a factor of $\sim3$ lower mean baseline flux. This comes from the simple fact that the two systems happen to be at slightly different viewing angles with respect to the CBD, and that the warped model blocks slightly more of the mass-trading emission. Finally, the warped model is the most strongly lensed model we consider because it is the most closely aligned; thus, the BH shadow width is at its maximum, as this is the angle at which the lens transits directly across the diameter of the background BH's shadow. 

\subsubsection{80\,$^{\circ}$ inclination}
\label{subsubsec:80deg}

\begin{figure}
    \centering
    \includegraphics[width=0.47\textwidth]{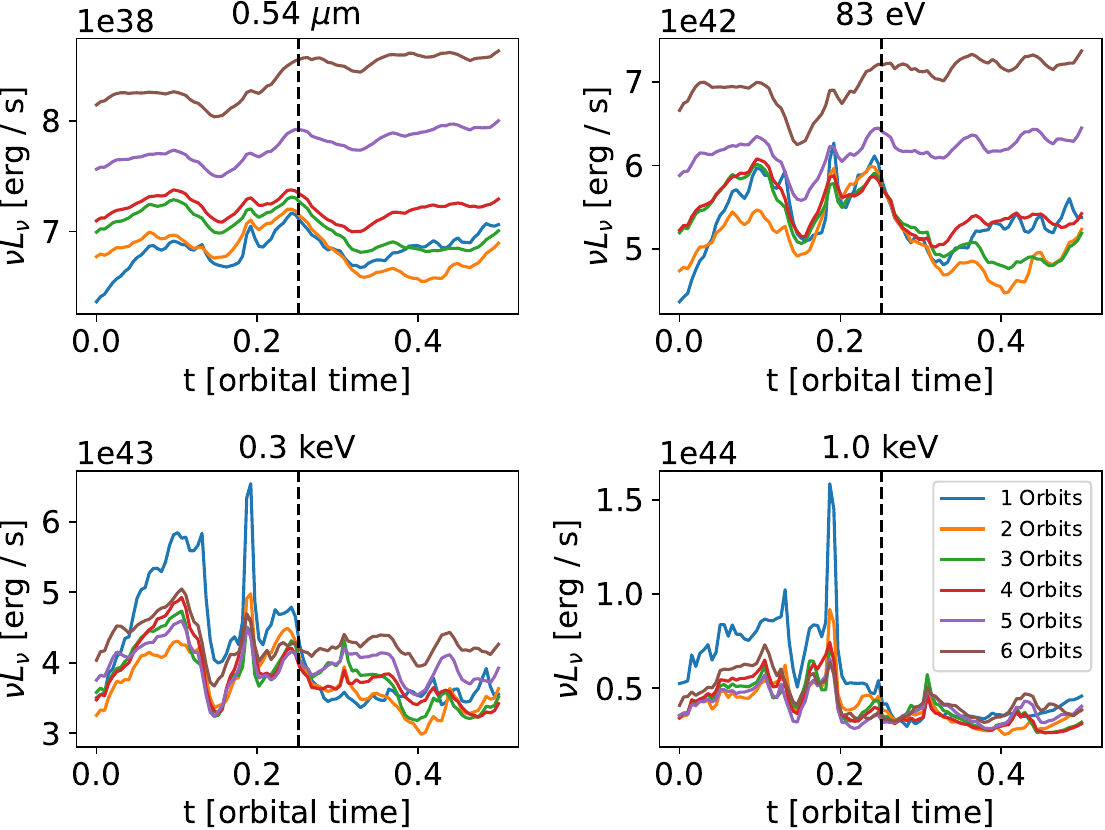}
    \caption{Light curves at the energies as labeled, phase-folded over an increasing number of 1-6 orbits, as indicated in the legend, in the $80^{\circ}$ inclination model. The vertical dotted black lines mark the moment of direct alignment to the line-of-sight and the gray regions around them represent the transtion duration of the BH shadow. Due to the less edge-on inclination, peaks caused by the mass-trading are now visible, which can obfuscate the SLF signature.}
    \label{fig:80deg_compare}
\end{figure}

We complete this section by returning to our fiducial model, without thinning or warping, but now examining the effects of having a much lower inclination angle. Rather than phase-folding all of the orbits, as in the previous sections, we choose to show the cumulative effect of phase-folding the light curve orbit by orbit, to highlight the effects of noise. This is shown in Figure~\ref{fig:80deg_compare}, showing light curves phase-folded over 1 to 6 orbits. At lower energy, the overall luminosity increases with the number of orbits folded, due to the lump orbiting into view, raising the baseline flux. This happens in all of our models, but its effect is small enough that phase-folding still enhances the SLF. Given enough orbits, lump-induced modulations of the baseline flux would phase-fold away. 

Potentially more consequential, is that without the CBD blocking much of the non-SLF emission, we are now privy to seeing the full range of EM emission mechanisms present in MBHBs, such as stream-stream and stream-cavity collisions, and perhaps the most significant is the emission coming from mass trading between minidisks. Looking at the light curves during the first orbit (blue) in Figure~\ref{fig:80deg_compare}, we see a mass-trading event occurs during this orbit. This presents as the spike to the left of the vertical dotted line, most extreme at higher photon energies. Without the CBD blocking the mass-trading emission, it dominates the signal at higher energies. However, we also see that the mass trading feature starts to wash away as we fold in more orbits. This is because the mass trading feature does not repeat on precisely the orbital timescale. If you compare the warped 2-orbit phase-folded curves in Figure~\ref{fig:fid_scale0p3_warp_compare} to the 2-orbit phase-folded curve in Figure~\ref{fig:80deg_compare}, you can see they are comparable, despite being different systems. But as we continue to fold in more orbits, the mass trading feature starts to vanish. It is our suspicion, that given a sufficient number of orbits to fold, this mass-trading feature may eventually average out entirely, leaving the SLF as the only clear signature. Unfortunately, we were unable to test this with our limited number of orbits. The practical complication of this is that if one observes a periodic signal truly originating from mass-trading, one might mistake this for the orbital period. It may be beneficial to phase-fold over more trial periods, to see if underlying patterns or false-positive SLF-mimicking signals could emerge.

It is worth noting another unique effect that was not present in the other models, either because they were too obscured by the CBD or they did not run for long enough. Although it is phase-folded away in Figure~\ref{fig:80deg_compare}, a very bright flare occurs during the fifth orbit, just before the streams between the minidisks collide (usually the brightest moment of the mass-trading event). Closer inspection revealed that as the minidisks began to trade mass, some of the material actually went right behind the foreground BH and got lensed. While the orientation of this shock-heating needs to be just right, this can be an additional source of increased flux during an orbit.

\section{Summary and conclusions}
\label{sec:DisCon}

We performed 2D hydrodynamical simulations of accreting binaries and coupled our findings to a general-relativistic ray-tacing code to characterize "self-lensing flares," and to asses their detectability.
We examined several models that also include parameterizing the disk scale height and warping of the CBD, for varying inclination angles between 80$^{\circ}$--90$^{\circ}$. We computed spectra at different snapshots, and analyzed light curves in four bands from the optical to X-rays, which we phase-folded for several orbits. To our knowledge, this is the first study of this kind. We built on previous work on lensed emission from binaries, which had adopted the microlensing approximation and simplified toy models for the intrinsic emission, to ascertain if SLFs can survive the more realistic accretion variability expected in these systems.

In our fiducial model, we saw that the spectra of SLFs are both harder and several standard deviations brighter, especially in the X-rays, compared to the average brightness during the rest of the orbit.  The fiducial model shows clear flares at all four energies we examined: $0.54~\mu$m, 83 eV, 0.3 keV and 1 keV. We also see a dip at the peak of the SLF, which is clearest at the lower energies where  obfuscating Doppler effects are less prominent. The SLF is highly chromatic, increasing with photon energy, unlike in the previous toy models. This is because the primary regions being lensed are the minidisks of whichever BH is in the background, which dominate the highest-energy emission. While detecting a SLF may be easier at higher energy, determining the shadow size of the BH from the dip may prove easier at softer bands.

We examined several model variants, which also exhibited SLFs with a dipping feature, but also made clear the importance of the inclination angle with respect to the CBD. When the viewing angle is high enough (i.e. sufficiently close to edge-on) that the CBD blocks a sizeable portion of the emission from the inner regions, the only emission from the gas in or near the individual BHs that can escape the system is associated with the SLF. This is because lensing bends this emission around the CBD blockage. However, when the viewing angle is low enough (far enough from edge-on) for mass-trading emission to become directly visible, its emission is both harder and brighter than even the SLFs.

In principle, given that the mass-trading happens quasi-periodically with frequencies slightly different than the orbital frequency, phase-folding at the orbital period for a sufficiently large number of orbits should average out the mass-trading contribution, recovering the SLF signature. Our hydrodynamic runs were of too short duration to test this definitively. If phase-folding does not eliminate the confusion with mass-trading flares, or enough data cannot be gathered to average the mass-trading over sufficiently many orbits, the viewing angle to unambiguously identify the SLFs may be limited to just a few of degrees above or below the CBD. While previous papers, such as Paper I and II, suggested viable viewing angles perhaps $4-5^{\circ}$ on either side from edge-on, this may be limited to closer to $1-3^{\circ}$ above the CBD, effectively cutting the number of potential SLF identifications by half. On the other hand, because the non-axisymmetric, possible warped CBD precesses, the viewing angle enabling observation of the SLF will evolve. One may therefore expect periodic SLFs would transition from visible to invisible over long CBD precession cycles. Furthermore, the lump orbits around the cavity wall on more observable timescales, on the order of several times the binary's orbital period. For certain viewing angles, it is possible the lump would temporarily block the SLFs while it is transits in front of the binary, creating an observable quasi-periodic SLF duty cycle.

While the focus of this paper is thermal emission, additional non-thermal emission mechanisms that may effect the SLF signature are possible, specifically those arising from magnetic fields, such as jet or coronal emission. During an SLF, if the foreground BH were to emit an optically thick jet it may geometrically block the lensed emission of the background BH. Or, if jet emission near the base of the background BH were to become lensed, the emission could change the shape of the SLF or potentially fill in the shadow-related dip. Additionally, unlensed emission from the jets of either BH will contribute to the light curve, which may have strong variations due to the beaming of the jet emission, and the variable viewing angle, which may or may not phase-fold away. Magnetic reconnection may also generate a non-thermal X-ray component. While this may not block the dip from the BH shadow because it is optically thin, it may still affect the overall signal. However, given again that this emission is chaotic, without periodicity, it is reasonable to think phase-folding may eliminate it. Regardless, 3D magnetohydrodynamic simulations would be valuable for clarifying the impact of non-thermal emission of the jets and coronae.

Although identifying the host galaxy of MBHBs with LISA prior to their merger may be difficult due to large sky localisation uncertainties (e.g. \cite{mangiagli2020}), triggered EM campaigns may be able to help \cite{lops2023a}. Ref.~\cite{Krauth2023} recently proposed to do so by searching for a source whose thermal X-ray emission abruptly disappears just prior to merger. SLFs provide yet another diagnostic for a compact binary. While light curves are noisy, the SLFs occur with a stringently periodic cadence when the binary orbit is evolving slowly, and our results here suggest that
phase-folding, even over just a few orbits, may suppress all other aperiodic accretion-based imprints to the EM signature, leaving the SLFs clearly identifiable. 

Finally, we emphasize that because SLFs and GWs are both intrinsically linked to the same orbital period,
and the SLFs furthermore arise at a known orbital phase (when the two BHs are aligned with the line of sight) they provide a novel and uniquely robust test for constraining the mass of the graviton by comparing the arrival times of the EM and GW signals. Furthermore, the identification of SLFs offers robust EM constraints on the orbital inclination and phase (and possibly eccentricity and nodal angle) of MBHBs detected by LISA, as well as a novel avenue for measuring black hole shadow sizes in systems that are beyond the resolving capabilities of existing VLBI facilities. 

\acknowledgments
We acknowledge support by NSF grant AST-2006176 (ZH) and NASA grant 80NSSC22K0822 (AM and ZH). JD acknowledges support by a joint Columbia/Flatiron Postdoctoral Fellowship. Research at the Flatiron Institute is supported by the Simons Foundation. This research was supported in part by the National Science Foundation under Grant No. NSF PHY-1748958. This research has made use of NASA's Astrophysics Data System.

{\it Software:} {\tt RAPTOR}, {\tt Sailfish}, {\tt python} \citep{travis2007,jarrod2011}, {\tt scipy} \citep{jones2001}, {\tt numpy} \citep{walt2011}, {\tt matplotlib} \citep{hunter2007}

%%%%%%%%%%%%%%%%%%%%%%%%%%%%%%%%%%%%%%%%%%%%%%%%%%
\section*{Data Availability}
The data underlying this article will be shared on reasonable request to the corresponding author.

\appendix
\section{Numerical validation}
\label{app-a}

We have previously tested and shown the numerical convergence of our work within \texttt{Sailfish} \citep[please see Appendix A of][]{Krauth2023}. For our work here we employed double the resolution of our previous work, which had already shown sufficient convergence. \texttt{RAPTOR} however, still required testing for our models.

The parameters of import are the number of pixels, the step size, and the AMR level. Pixels defines the number of pixels in our camera viewing window, horizontally and vertically (higher is more refined). Step-size defines how large of a step we take from the camera to the source (lower is more refined). The AMR level defines how many subgrids the refinement can break down into. For example, if there is a sufficiently large gradient in intensity between to neighboring pixels, refinement of the camera pixels may occur (larger number of levels means more refinement). When treating the step size for all models, we incorporated a relation to decrease the step size in proportion to the vertical disk height of the source of emission being traced, to try to ensure we do not over-step the emitting pixel.

\begin{figure}
    \centering
    \includegraphics[width=0.47\textwidth]{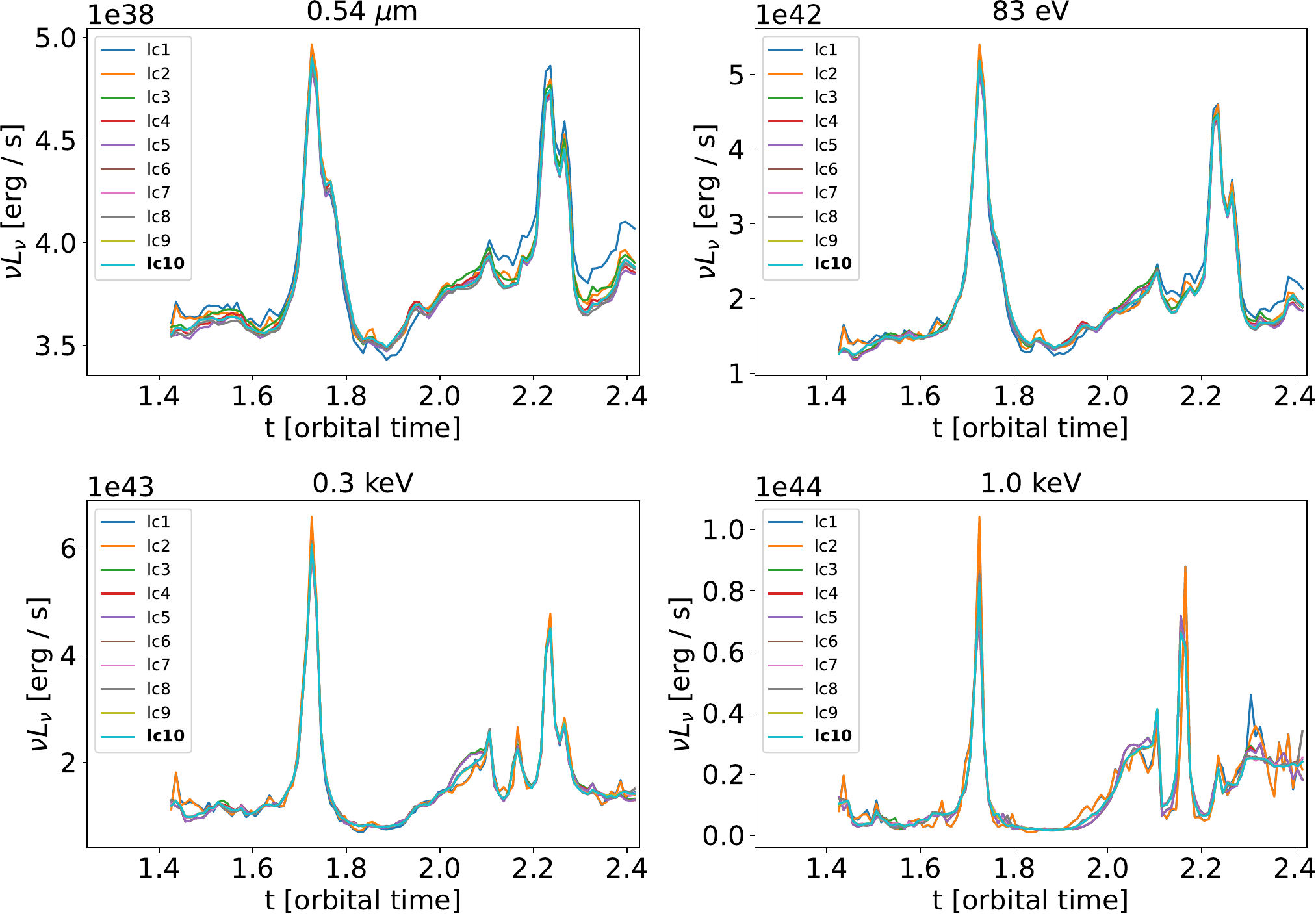}
    \caption{Convergence test for several different model with the parameters for each described in Table~\ref{tab:ResTet}. The final model chosen for the study is the lc10, listed in bold.}
    \label{fig:convtest}
\end{figure}

Table~\ref{tab:ResTet} shows the parameters we modified in order to attain convergence, and Figure~\ref{fig:convtest} show the corresponding light curves for 100 checkpoints of an arbitrary orbit we used to analyze. While most curves past the second or third model look fairly comparable, there are moments, especially in the higher energies, that still fluctuated. Thus, we wanted to use a resolution high enough that we could trust the convergence of values at all times, not just the well-behaved times. We gradually tried each of the parameters individually, and in tandem, until a significant difference could no longer be seen. Although model lc9 would have actually been the highest resolution, the computational time was exorbitant. Although there was not a significant difference between models lc6 and lc9, with the increased AMR level we decided the best way to ensure consistency was to split the difference and create a grid that had an AMR of 3 generically, but was pre-set to AMR 4 when within the Einstein radius of either BH. This reduced the run-time to something manageable while ensuring we still had the highest resolution and convergence within the important regions of interest that are most likely to need it. The designation of this optimal model is lc10.

\begin{table}
\caption{Summary of resolution studies we performed to ensure convergence of ray-tracing results. The parameters we employ in this work are given by lc10, indicated in boldface.}
\centering
\setlength{\tabcolsep}{10pt}
\begin{tabular}{l| c c c}
 \hline \hline
 Model &Pixels &Step size &AMR \\
 \hline
 lc1 &200x200 &0.04 &1 \\
 lc2 &200x200 &0.02 &1 \\
 lc3 &200x200 &0.02 &2 \\
 lc4 &200x200 &0.01 &2 \\
 lc5 &200x200 &0.005 &2 \\
 lc6 &200x200 &0.01 &3 \\
 lc7 &300x300 &0.01 &3 \\
 lc8 &200x200 &0.005 &3 \\
 lc9 &200x200 &0.01 &4 \\
 \textbf{lc10} &200x200 &0.01 &3/4 \\
 
 \hline \hline
\end{tabular}
\label{tab:ResTet}
\end{table}

\end{document}